# Mobile Agent based Market Basket Analysis on Cloud

Vijayata Waghmare
Department of Information Technology
Maharashtra Institute of Technology
Pune, India
vijayataw@gmail.com

Debajyoti Mukhopadhyay
Department of Information Technology
Maharashtra Institute of Technology
Pune, India
debajyoti.mukhopadhyay@gmail.com

*Abstract*— **This paper describes the design and development of a location-based mobile shopping application for bakery product shops. Whole application is deployed on cloud. The three-tier architecture consists of, front-end, middle-ware and back-end. The front-end level is a location-based mobile shopping application for android mobile devices, for purchasing bakery products of nearby places. Front-end level also displays association among the purchased products. The middle-ware level provides a web service to generate JSON (JavaScript Object Notation) output from the relational database. It exchanges information and data between mobile application and servers in cloud. The back-end level provides the Apache Tomcat Web server and MySQL database. The application also uses the Google Cloud Messaging (GCM) for generating and sending notification of orders to shopkeeper.**

*Keywords- mobile cloud computing, association rule mining, location-based services, JSON*

## I. Introduction

Mobile shopping involves mobile devices as electronic gadget. Development in smart phones results in the increased rate in its usage and information anytime anywhere has become style mantra. But mobile devices have limited memory and processing speed. Cloud in other hand provides large storage and speed for heavy data. Increased need of processors and storage devices results in the development of cloud computing. Mobile cloud computing is the area where three entities are involved: Mobile devices, communication network and cloud as a server. Data storage and data processing occurs outside the mobile devices i.e. on cloud and results are displayed through output devices like screen or speakers. GPRS, Gmail, and Google Maps are already being used are pioneer examples of mobile cloud computing. Thus, mobile cloud computing overcomes the weaknesses of mobile devices like, short storage area, and processing power.

This research covers these weaknesses and implements association rule mining on the data collected from mobile application. This application is specially designed for bakery product purchasing in the city nearby the customer location. It uses Wi-Fi and cellular network to get current position of the customer and displays any registered bakery shop on map within 5km area from customer's location. Association rule mining as a method of data mining is used to find the top associated famous products purchased from chosen bakery. Technically, data mining is the process of finding correlations or patterns among dozens of fields in large relational databases. Association rule mining is widely used in market basket analysis. This process benefits retailers in several ways for marketing or planning shelf space.

A location-based mobile application for bakery product shopping was designed and developed to find nearby bakery shops, association among the products purchased from bakery, display association to customer side screen, post order, and it is deployed on cloud. The combination of web map service API and association rule mining using mobile on cloud, it is possible to collect large scale shopping habit of people, with lower data collection cost.

The computing process solves the problem of limited computing power of mobile devices. This framework uses best third-party libraries to avoid unnecessary processes. The requests are arranged in queue for the confirmation of order by the shopkeeper. Web services are used to manage the connection between front-end and back-end.

The Framework presented in Figure 1 is the working principle of this application.

Large amount of data collection will be possible using mobile as users can place order from anywhere and service provision will be easier. Just few clicks and order will be placed. Also, both customers as well as shopkeepers can take the benefit of mined information for their own benefit and also for marketing the products.

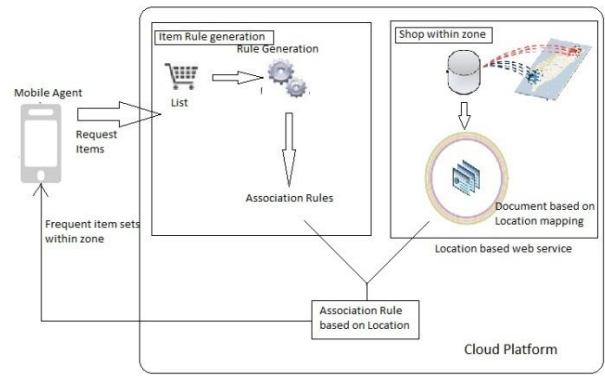

Figure 1. Proposed Framework

The paper is organized as follows; section 2 presents the related work, which includes three main technologies and concepts. Section 3 is all about design and development of the application and the paper is concluded in section 4.

## II. RELATED WORK

The main advantage of using cloud is to increase speed and performance. The combinations of Location-based services and Association rule mining gives best results in the form of Graphical User Interface (GUI). This section discusses the existing projects related to these techniques.

### A. Mobile Cloud Computing

Mobile cloud computing is the area where three entities are involved: Mobile devices, communication network and cloud as a server.

In mobile cloud computing resources are stable but application may move. The large applications can be decomposed to smaller ones to process concurrently. This approach is called as application partition. Offloading is the process of transmitting mobile application on cloud. This saves the device memory, processing power and ultimately battery consumption [1].

The typical services needed by a mobile cloud client are, synchronization, push i.e. updates the notifications sent by the cloud server, offlineApp automatically handles synchronization and notification, network, database, interAppBus provides low level coordination between applications [2].

### B. Association Rule Mining

In this paper [3], author proposes a framework the attempts to reduce the communication overhead in existing mobile agent based distributed association rule mining. MAD-ARM is the Mobile Agent based distributed data mining framework. It consists of knowledge server which works on the generation of association rule and data coming are from different remote sites. The itemsets are always updating on remote sites at the stationary agent.

The authors of this paper [4], represented a scenario for data analysis. This is clearly based on the environment where data stream mining process runs on users smart mobile phones. As the data streams in continuously, possible concept drift is updated. There is one central mobile decision agent which controls several others stream mining agents. Stream mining agents working on local handsets decides the best possible algorithm to run on local data. Algorithm is chosen dynamically at each handset.

### C. Location-based services

Any product, service or application that uses the location information of mobile subscriber is called as location based service. Location based services uses the latitude and longitude information.

A context-based multimedia content management system (MCMS), whose various types of contents are easily gathered from everywhere at anytime using mobile phones, and stored in a web server as a multimedia database [5].

[6] Describes a location based text mining approach to classify texts into various categories based on their geospatial features, with the aims to discovering relationships between documents and zones. There are three main components in this framework, including geographic data collection and reprocessing, mapping documents into corresponding zones, and framing maximize zones. Data extraction and processing is takes place based on zones.

Tourism industry has also takes the advantage of location based services. This application is designed and developed using cloud based platform. It finds out the location of tourists, where they are heading or looking. This is possible by using the digital compass. It also calculates distance between current position and places, it shows nearby attractions, and provides direction. It was built on Amazon Web Services cloud platform [7].

## III. DESIGN AND DEVELOPMENT

The design and development of the application is a three level process, Front-end, middle-ware level and back-end level. Each level uses different technologies and work in collaboration with each other.

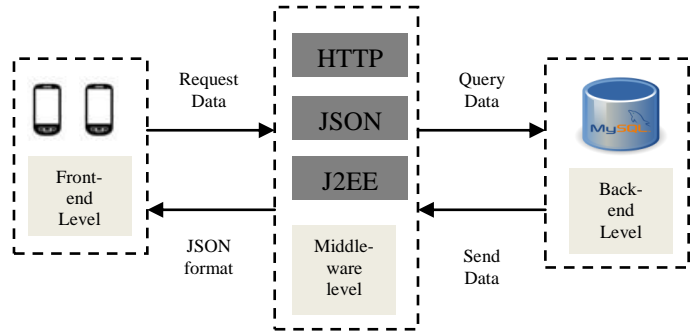

Figure 2. Data transfer by web services between three-tiers

As shown in Figure 2, middleware-ware processes the request from front-end and query to the database located at the back-end, after reading data from database it performs operations that generates JSON format file and allows front-end to parse it. Working of each level is described as follows:

### A. Front-end

An application is designed and developed for android platforms to be able to run on android supported devices such as HTC Desire C/V/Z/S, Samsung Galaxy Grand 2/Y/Nexus, Sony Xperia series, etc [8]. for creating an application for these devices developers uses Eclipse (Android Development Tools) ADT bundle which includes android SDK tools, and version of Eclipse Integrated Development Environment (IDE) with build-in ADT.

Registration for both customers and shopkeepers are available and they can login from their respective devices.

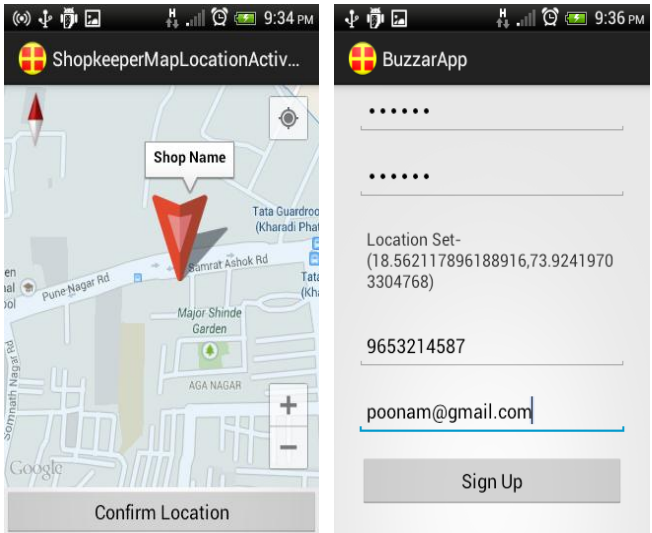

Figure 3. (a) Displaying selection screen of location of shop. (b) Displaying co-ordinates in the form of latitude and longitude

In the shopkeeper side front-end, i.e. when login activity is done by shopkeeper, current location or the any location where shop is located is selected from map. As shown in Figure 3(a), the application displays current location on map. The red arrow in the map gives the selected location while the circle in the upper right corner gives the current location. It is recommended that bakers must register from the shop itself but even if they can't they can choose any location by moving map and tapping desired location. This map is obtained by using Google Map Service API (Application Program Interface) [9]. In this view first is the map and second one i.e. Figure 3(b) is the accepted co-ordinates of the location of shop from map in the form of latitude and longitude.

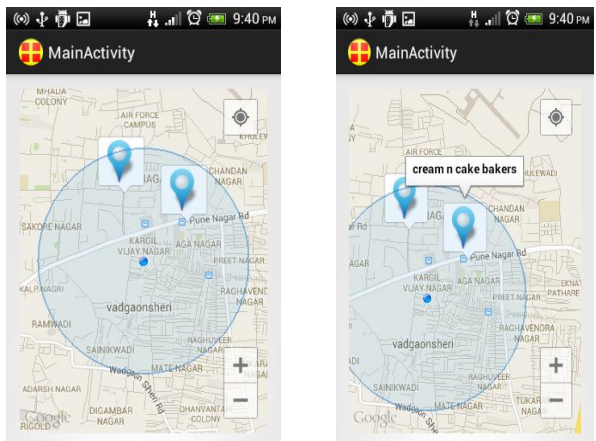

Figure 4. (a) Displaying registered bakery shops within 5(km) on a map. (b) Displaying annotation of selected shop

As shown in Figure 4, the application displays nearest bakery shops on a map within 5 km radius from the customer's current location. Here map and annotation of the selected shop is shown. Figure 4(a), shows the nearby shops from the customer's location which falls within 5 km area. Area is highlighted as blue circle. In figure 4(b), when customer taps any annotation for choosing a shop, the name of the shop appear and when clicked on the name the list of best associated product is displayed which were obtained from implementing apriori algorithm on server side data stored in MySQL database. This screen further leads to the screen, displaying products available in the shop.

Google play services [10] library provides Google API for connecting and making call to any other Google play services. An internet connection needs to be added in the Android Manifest file and obtained a key to started map service. While shopkeeper registration location is saved to the MySQL database permanently so that customer's will get the qualitative and accurate location-based results.

At the shopkeeper side when login in bakers can add new items to the product list from the item list already provided and can also add its price. So, different shops can have different prices for the same item.

The home screen of Shopkeeper is as shown in Figure 5(a), from this screen shopkeeper controls all the activities. Apart from adding new products, it performs two more activities, one is handling pending request and other one is finding association rule. When the request is sent by customer, shopkeeper gets the notification. All the unhandled notifications are kept in a queue as shown in Figure 5(b). Notifications are generated using Google Cloud Messaging (GCM) [11]. GCM connection servers take message from our server and sends these messages to the application which is GCM enabled, connection server are provided for HTTP.

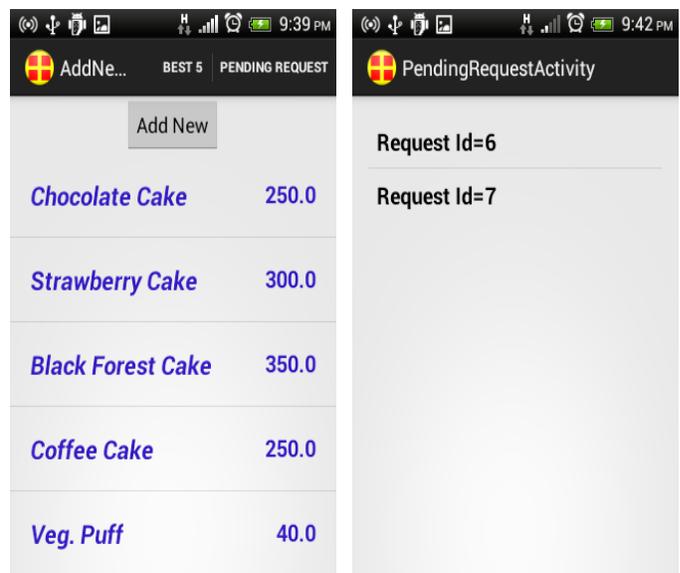

Figure 5. (a) Shopkeeper Home Screen. (b) Pending Request Activity

After receiving the request notification from customer, bakers can confirm or decline request, declined request has been deleted from transaction and apology message can be sent to customer via, Gmail, Google drive, or other mail provider. If the request is confirmed a confirmation message is sent to customer by above mentioned mail providers along with billing.

Next very important part of the project is Association rule mining which is popularly known as market basket analysis. After the location and shop is selected by customer best five combinations which has highest frequency of purchases are shown on screen which ultimately gives recommendations to the customers. Association between different items is the frequent itemset mining; it discovers interesting correlation relationship among huge amount of data. The benefits of finding such patterns are useful in catalog design, cross-marketing, and customers shopping behavior analysis. Association rule is the representation of pattern that reflects items that are purchased together. This interestingness among items is calculated using two measures, support and confidence. Support considers all possibilities that are recorded while confidence is the guarantee that those possibilities exists. The best rule is the rule that satisfy minimum support threshold and minimum confidence threshold value. The recommendations obtained from purchasing of bakery products are as shown in Figure 6. Apriori algorithm is used to obtain these combinations.

If we say that, A and B are two items or itemsets then,
$$support(A=>B) = P(A \cup B)$$
And confidence is,
$$confidence(A=>B) = P(B|A)$$
$$= support(A \cup B) / support(A)$$
$$= support\_count(A \cup B) / support\_count(A)$$

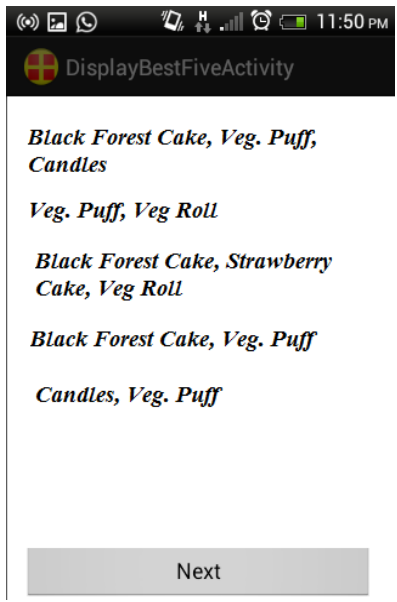

Figure 6. Recommendations based on previous records.

WEKA [12] is used specifically for performing data mining tasks. It is written in java. For data mining through java programming, .jar file of weka api is imported in the project. Weka consists of ARFF files which are external representation of an instances class.

*B. Middle-ware*

Data exchange between front-end and back-end level database of bakery product shopping application happen via middleware. Android application sends http request and web services in the middle-ware sends query to fetch data from MySQL database, JSON [13], output is generated from the obtained data from database which is parsed at the front-end to display appropriate output. JSON is language independent so it is useful for data exchange, storage and communication by any programming language. Web services in JSON format makes data access stored in MySQL database easy. As shown in figure 1, JSON works on four steps for communication. At first step application sends http request to the web services, web services using JSON format accept the data. In second step a response is generated by back-end level and middle-ware level which is in the form of JSON output. The query is sent to the back-end database to obtained data from relational MySQL database.

In third step, JSON response is generated by web service which is sent to the application side and read it. JSON object is needed to be decoded into string which can be displayed on screen. In step four, these objects are ready to be displayed on screen.

Parsing data using JSON parser is quite easier than parsing data using XML Parser. JSON provides data in the form of objects whenever it is reading from source, while Object for JSON is created at the time of transmission of data over network. Using JSON is the right choice when it comes to mobile communication as it does not have a specific tag format, which avoids the consumption of bandwidth. So instead of transmitting and getting back XML file Communication via JSON objects is appropriate.

Java Platform Enterprise Edition extends the Java Platform Standard Edition. It provides an API for multitier applications and web services. It also provide runtime environment for developing and running enterprise software. Java Database Connectivity (JDBC) is the standard interface for Java. Without requiring a graphical user interface Java Servlet API can enhance consistency for developers.

*C. Back-end*

In back-end level the server are built as shown in Figure 7, infrastructure consists of the cloud based net4india [14] cloud platform. Virtual Private Server (VPS) of net4india is used as a cloud to run the apache tomcat web server. This is available in resizable compute capacity. Relational database services are provided which are used for having a MySQL database server in the cloud to host. Transaction data is fetched to different text file format for performing association rule mining. the rules which has confidence more than 60% are calculated and top 5 association rules are generated which are allowed to be display on front-end of client side. These rules are accessed using JSON objects and results are transmitted. Apache tomcat is a servlet container and open-source web server.

The performance of the algorithm is influenced by the dimension of the data set and also by the support factor [15]. The first part of algorithm finds all frequent itemsets while the second part generates the strong association rule.

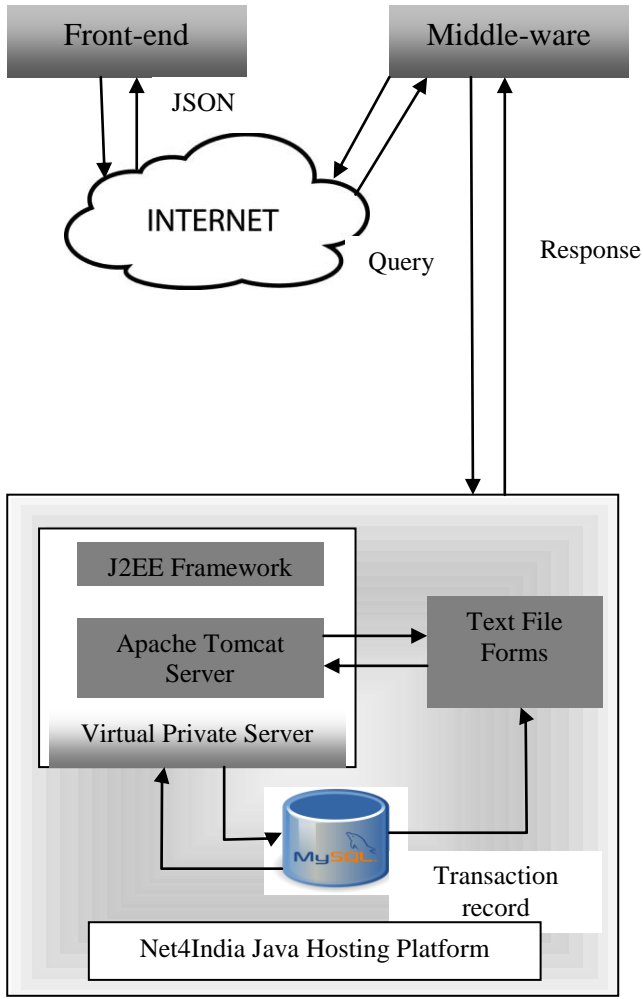

Figure 7. Architecture at back-end level

## IV. CONCLUSION

We designed and developed a location-based mobile shopping application for bakery shops for android platform at the front-end. This application displays nearby bakery shops that are registered to the application. Data exchange between different levels of architecture are operated using web services and that generated JSON format for data transfer. The server is built in cloud by using net4india hosting services.

With the help of mobile cloud computing mobile processing and storage is transferred to cloud as a server, which helps in saving battery consumption and improves the performance or speed of execution. Use of the location-based services gives flexibility and attractive looks to the application. No extra charges are applied; application can be downloaded and utilized in regular data charges.

Relationship between products gives the attraction information of related bakery. Information about frequently purchased bakery products can help in cross marketing. The design of the project is in such a way that application can be used for any shop system we just need to change the name of products in the coding.

In future works we consider using more automation in the application by providing information without registration process and whole transaction will happen on the mobile number provided. We can integrate the routing of GPS to provide direction and distance measurement between shop and customer.